\documentstyle[aps,twocolumn]{revtex}

\begin{document}
\author{V.Yu.Irkhin$^{*}$ and Yu.P.Irkhin}
\address{Institute of Metal Physics, 620219 Ekaterinburg, Russia}
\title{Charge screening and magnetic anisotropy in metallic rare-earth systems}
\maketitle

\begin{abstract}
The calculation of magnetic anisotropy constants is performed beyond the
point charge model for a continuous charge density distribution of screening
conduction electrons. An important role of the non-uniform electron density,
in particular, of the Friedel oscillations, in the formation of crystal
field is demonstrated. Such effects can modify strongly the effective ion
(impurity) charge and even change its sign. This enables one to justify the
anion model, which is often used at discussing experimental data on
hydrogen-containing systems. Possible applications to the pure rare-earth
metals and RCo$_5$ compounds are discussed. The deformation of magnetic
structure near the interstitial positive muon owing to the strong local
anisotropy, and the corresponding contribution to the dipole field at the
muon are considered.
\end{abstract}


The old problem of strong magnetic anisotropy in the rare-earth-based
intermetallic systems\cite{irk}, which has a great practical importance, is
up to now extensively investigated. Recently the anisotropy induced by
interstitial hydrogen atoms\cite{wal,mush} and positive muons\cite{ii} has
been discussed.

It is accepted now that the main mechanism of magnetic anisotropy origin in
the rare-earth systems is the crystal field one. Estimations of the
anisotropy constants are usually performed in the point-charge model. At the
same time, this model leads frequently to difficulties and contradictions
with experimental data. For example, the calculated anisotropy constant $K_1$
in RCo$_5$ compounds turns out to be very large and have an incorrect sign%
\cite{ros}. Thus screening of the crystal field should play an important
role. This screening is often taken into account by introducing the
effective ion charge $Q^{*}$ which can differ considerably from the bare ion
charge. In particular, the fitted charge of hydrogen ions for the RCo$_2$%
-based systems turns out to be negative (the anion model\cite{wal,mush}).
Physical explanation of such situations is not simple. Shielding of crystal
fields in ionic solids was treated in Ref.\cite{stern}. Screening by
conduction electrons, which form some localized levels near the rare earth
ion, was discussed in the paper\cite{ros}. However, these effects turn out
to be insufficient for explaining experimental data.

In the present paper we perform calculations of magnetic anisotropy
constants with account of continuous charge-density distribution in a metal.
We shall demonstrate that the magnetic anisotropy is strongly influenced not
only by the total charge induced by surrounding ions or impurity centres,
but also by a concrete form of screening electron density, in particular, by
the Friedel oscillations.

We consider the magnetic ion at the point ${\bf r}=0$ in the crystal field
of the surrounding charges. The spherically symmetric potential $V_{cf},$
which is induced by a center at the point ${\bf r=R}$ and originates from
the point charge $Q_0$ and conduction electron charge density $Z(|{\bf r-R|}%
),$ has the form 
\begin{equation}
V_{cf}({\bf r})=\frac{Q_0+Q_{el}(|{\bf r-R}|)}{|{\bf r-R}|}  \label{V}
\end{equation}
where 
\begin{equation}
Q_{el}(r)=4\pi \int_0^r\rho ^2d\rho Z(\rho )  \label{qq}
\end{equation}
is the conduction electron charge inside the sphere with the radius $r,$ $%
Q_{el}^{\prime }(r)=4\pi r^2Z(r)$, the system of units where the electron
charge $e=-1$ is used. At large $r$ complete screening take place, so that $%
Q_{el}(\infty )=-Q_0$.

The anisotropy constants are determined from the angle dependence of the
energy of the magnetic ions in the crystal field 
\begin{equation}
\delta {\cal E}_{cf}=-K_1\cos ^2\theta +...  \label{an}
\end{equation}
First we discuss the point charge model. The dependence (\ref{an}) is
calculated with the use of the expansion in $r,$ which corresponds to the
expansion in small radius of $f$-shell:
\begin{equation}
\frac 1{|{\bf r-R}|}=\frac 1R\left[ 1+\frac rR\cos \widetilde{\theta }+\frac{%
r^2}{2R^2}(3\cos ^2\widetilde{\theta }-1)+...\right]   \label{rr}
\end{equation}
Here $\widetilde{\theta }$ is the angle between the vectors {\bf R} and {\bf %
r, }which can be expressed in the spherical coordinate system of the crystal
as 
\begin{equation}
\cos \widetilde{\theta }=\cos \theta \cos \theta _{{\bf R}}-\sin \theta \sin
\theta _{{\bf R}}\cos (\phi -\phi _{{\bf R}})  \label{tet}
\end{equation}
where $\theta _{{\bf R}}$ and $\phi _{{\bf R}}$ are the polar and azimuthal
angles of the vector ${\bf R}${\bf . }Substituting (\ref{tet}) into (\ref{rr}%
) and picking out the term, which is proportional to cos$^2\theta ,$ one
obtains for the point-charge contribution (cf. Refs.\cite{kas,mush}) 
\begin{equation}
K_1^{pc}=-3\Lambda \langle r_f^2\rangle \alpha _JJ(J-1/2)Q_0e^2  \label{pc}
\end{equation}
where $\langle r_f^2\rangle $ is the average square of the $f$-shell radius, 
$J$ is the total angular momentum of rare-earth ions, $\alpha _J$ is the
Stevens factor, 
\begin{equation}
\Lambda =\sum_{{\bf R}}\frac{3\cos ^2\theta _{{\bf R}}-1}{R^3}
\end{equation}
For the hcp lattice with the parameters $c$ and $a$ (pure rare earth metals)
we derive after summation over nearest-neighbor magnetic ions the standard
result (see Ref.\cite{ros}) 
\begin{equation}
\Lambda =6a^{-3}\left( 16\frac{y^2-2/3}{(4/3+y^2)^{5/2}}-1\right) \simeq
2.44a^{-3}(\sqrt{8/3}-y)  \label{ree}
\end{equation}
where $y=c/a\simeq 1.57\div 1.59$ for heavy rare earths\cite{Tay}. The small
geometrical factor $\sqrt{8/3}-y=1.633-y\sim 0.05$ occurs due to that the
contributions from the magnetic ions in the same plane and in neighbor
planes have opposite signs and almost cancel each other.

For the RCo$_5$ compounds, where $R=a/\sqrt{3}$ for six neighbor Co ions in
the same plane and $R=\frac 12\sqrt{a^2+c^2}$ for six Co ions in the upper
and lower planes, we have\cite{ros} 
\begin{equation}
\Lambda =6a^{-3}\left( 16\frac{2y^2-1}{(1+y^2)^{5/2}}-3^{3/2}\right) \simeq -%
\frac{23.4}{a^3}  \label{co}
\end{equation}
with $y=c/a\simeq 0.8,\,a\simeq 5$\AA $.$ In this case the cancellation of
contributions from different planes does not play a crucial role, so that
the expression in the brackets is never small.

Consider the case where anisotropy is induced by random next-neighbor
substitutional or interstitial impurities. For an impurity in the octahedral
interstitial of the hcp lattice\cite{kron} we have $R=a/\sqrt{2},$ $\cos
\theta _{{\bf R}}=1/\sqrt{3}.$ Then $\Lambda =0$ for the ideal hcp lattice,
and the impurity contribution to $K_1$ vanishes, as well as (\ref{ree})
(note that the situation is different in the hydrogen-containing RCo$_2$
systems\cite{mush}). However, there occurs the local anisotropy term 
\begin{equation}
\delta {\cal E}_{cf}=-K_{1\text{loc}}\cos ^2\widetilde{\theta }  \label{anl}
\end{equation}
with 
\begin{equation}
K_{1\text{loc}}^{pc}=12\sqrt{2}\langle r_f^2\rangle \alpha
_JJ(J-1/2)Q_0e^2/a^3.
\end{equation}
Note that for the positive charges $Q_0$ the signs of $K_{1\text{loc}}$ and $%
K_1$ coincide.

Now we treat the case of a continuous charge distribution of the screening
conduction electrons. Similar to (\ref{rr}), the potential (\ref{V}) can be
also expanded at small $r$. Performing the expansion of the integral (\ref
{qq}) in
\[
|{\bf r-R}|-R=-r\cos \widetilde{\theta }+(r^2/2R)\sin ^2\widetilde{%
\theta }+...
\]
we obtain up to $r^2$ 
\begin{eqnarray}
Q_{el}(|{\bf r-R}|) &=&Q_{el}(R)-4\pi R^2Z(R)r\cos \widetilde{\theta }+2\pi
Rr^2  \nonumber \\
&&\ \ \ \ \times \{Z(R)+[Z(R)+RZ^{\prime }(R)]\cos ^2\widetilde{\theta }\}
\end{eqnarray}
Taking into account the expansion (\ref{rr}) we have  
\begin{eqnarray}
&&\ \ \frac{Q_{el}(|{\bf r-R}|)}{|{\bf r-R}|}
\begin{tabular}{l}
=
\end{tabular}
\frac{Q_{el}(R)}R[1-\frac rR\cos \widetilde{\theta }+\frac{r^2}{2R^2}(3\cos
^2\widetilde{\theta }-1)]  \nonumber \\
&&\ \ \ -2\pi [Z(R)(2rR\cos \widetilde{\theta }-r^2\sin ^2\widetilde{\theta }%
)-RZ^{\prime }(R)r^2\cos ^2\widetilde{\theta }]
\end{eqnarray}
Then we derive 
\begin{equation}
K_1/K_1^{pc}=Q^{*}/Q_0  \label{rat}
\end{equation}
with the effective charge 
\begin{equation}
Q^{*}=Q_0+Q_{el}(R)-\frac 43\pi R^3[Z(R)-RZ^{\prime }(R)]  \label{qef}
\end{equation}
Thus we have obtained the expression for the effective charge $Q^{*},$ which
should be used at calculating the observable anisotropy constant $K_1$. We
see that the ratio (\ref{rat}) depends explicitly, besides the total charge
inside the sphere with the radius $R,$ also on the charge density $Z(R)$ and
the derivative $Z^{\prime }(R).$ The latter quantity can be large provided
that the rare-earth ion lies in the region where the charge density changes
sharply. Note that for the constant charge density $Z$ ($Z^{\prime }(R)=0,$ $%
Q_{el}(R)=\frac 43\pi R^3Z$) we have $K_1=K_1^{pc}$ and screening is absent
(this is connected with that the charge in the sphere increases as $R^3,$
but the quadrupole interaction decreases as $R^{-3}$).

To obtain the value of $Q^{*},$ one has to investigate the charge screening
for a concrete electronic spectrum, which is, generally speaking, a very
difficult task. We discuss the one-centre screening problem within a simple
model of free conduction electrons ($E=k^2/2$) in the impurity-induced
rectangular potential well which has the width $\ d$ and depth $E_0=k_0^2/2$%
\cite{dan}. This model enables one to calculate the charge distribution of
screening conduction electrons in terms of the scattering phase shifts $\eta
_l$. The value of $k_0$ should be determined for given $k_F$ and$\ d$ from
the Friedel sum rule 
\begin{equation}
Q_0=\frac 2\pi \sum_{l=0}^\infty (2l+1)\eta _l(k_F)  \label{fri}
\end{equation}
The phase shifts are calculated as 
\begin{eqnarray}
\tan \eta _0(k) &=&\frac{kd\,j_1(kd)-\beta _0j_0(kd)\,}{kd\,n_1(kd)-\beta
_0n_0(kd)},\,\beta _0=\widetilde{k}d\frac{j_1(\widetilde{k}d)}{j_0(%
\widetilde{k}d)} \\
\tan \eta _{l>0}(k) &=&\frac{kd\,j_{l-1}(kd)-\beta _lj_l(kd)\,}{%
kd\,n_{l-1}(kd)-\beta _ln_l(kd)},\,\beta _l=\widetilde{k}d\frac{j_{l-1}(%
\widetilde{k}d)}{j_l(\widetilde{k}d)}
\end{eqnarray}
where $j_l(x)$ and $n_l(x)$ are the spherical Bessel and Neumann functions, $%
\widetilde{k}^2=k_0^2+k^2.$ The disturbance of the charge density is given
by 
\begin{equation}
\delta Z(r)=-\int_0^\infty kdk\,\delta \rho (k,r)
\end{equation}
For $r>\ d$ one obtains 
\begin{eqnarray}
\delta \rho (k,r)/\rho _0(k) &=&\sum_l(2l+1)\{[n_l^2(kr)-j_l^2(kr)]\sin
^2\eta _l  \nonumber \\
&&\ \ \ \ -j_l(kr)n_l(kr)\sin 2\eta _l\}
\end{eqnarray}
with $\rho _0(k)=k/\pi ^2.$ For $r<\ d,$ joining of the wavefunctions at the
boundary of the potential well $r=\ d$ yields 
\begin{eqnarray}
&&\ \ \ \ \delta \rho (k,r)/\rho _0(k) 
\begin{tabular}{l}
=
\end{tabular}
\sum_l(2l+1)[j_l(kd)\cos \eta _l  \nonumber \\
&&\ \ \ \ -n_l(kd)\sin \eta _l]^2[j_l(\widetilde{k}r)/j_l(\widetilde{k}%
d)]^2-1
\end{eqnarray}

The parameter $d$ should be determined by the geometry of the lattice near
the impurity. In Ref.\cite{dan}, where impurities in the Ag host were
considered, $d$ was chosen to be equal to the Wigner-Seitz radius, so that $%
k_Fd=2$. In the case where the impurity is localized in an interstitial the
choice of $d$ may be different.

We have performed the calculations for $k_Fd=2$ and $k_Fd=3.$ At $Q_0=1$ Eq.(%
\ref{fri}) yields $k_0d=1.46$ and $k_0d=1.235$ respectively. With increasing 
$k_F,$ the number of $l$ values to be taken into account in the sums grows;
for $k_Fd=3$ the contributions up to $l=3$ are appreciable. The results are
presented in Figs.1-3. Note that a weak singularity occurs at the joining
point $r=d$.

One can see that at $R<d$, except for the case of very small $R$ where $%
Q^{*}(R)$ slightly decreases, the derivative term results in that $Q^{*}(R)$
grows (despite an increase of $|Q_{el}(R)|$). For $R\simeq \ d,$ where $%
Z^{\prime }$ is maximum, the non-uniform distribution of electron density
leads to that the effective ion charge $Q^{*}$ is positive and exceeds
considerably its bare value $Q_0$ ($Q_0=1$ in our case). At the same time,
with further increasing $R$ the situation changes drastically: $Z^{\prime }$
decreases and becomes negative, so that ``overscreening'' of the ion charge
occurs. Such a situation corresponds to the ``anion model''. In the simple
model under consideration, the values about $-0.5$ can be obtained for $%
Q^{*} $ (the value of -1 was assumed for the hydrogen ions in Ref.\cite{mush}%
). At large distances $Q^{*}$ tends to zero, but considerable oscillations
of the effective charge sign take place, which attenuate rather slowly. One
can see from Fig.2 that the contribution of the derivative term in (\ref{qef}%
) predominates at large $R$ too. Of course, more complicated models are
required for a regular lattice of screened charges, but the conclusion about
an important role of the non-uniform distribution of electron charge should
hold also in that case.

The effects of screening under consideration can be important for the values
of the anisotropy constants in both stoichiometric rare-earth compounds and
systems containing hydrogen and other impurities. The experimental data on
the first magnetic anisotropy constant of the heavy rare-earth metals and
corresponding RCo$_{5+x}$ compounds at low temperatures are presented in
Table 1. The values of the effective charge $Q^{*}$ are calculated by using (%
\ref{rat}), (\ref{pc}). One can see that the values of $Q^{*}$ for the pure
rare earths are of order of unity. At the same time, the effective charges
of the Co ions in the RCo$_{5+x}$ systems turn out to be negative and very
small (a similar situation takes place in the case where R is a light rare
earth\cite{ros}), so that a weak ``overscreening'' occurs. Note that in fact
the effective charges may be different for two kinds of the Co ions which
contribute (\ref{co}), since the corresponding values of $R$ are different.
Besides that, a possible role of the neighbor rare-earth ions in the crystal
field formation should be considered in such a situation.

Table 1. The total angular momenta $J$, Stevens factors $\alpha _J,$ average
squares of the $f$-shell radius $\langle r_f^2\rangle $ (atomic units) and $%
c/a$ ratios for rare-earth elements; the experimental values of $K_1$
(K/rare-earth ion) for the pure heavy rare earths\cite{irk} and RCo$_{5+x}$
systems (the contribution of rare-earth sublattice\cite{Erm}), and the
corresponding calculated values of $Q^{*}.$ The values of $K_1^{\exp }$ for
Er and Tm are obtained from the anisotropy of magnetic susceptibility (see
Ref.\cite{Coq}). The values of the local anisotropy constant $K_{1\text{loc}%
} $ for an impurity in the octahedral interstitial of a rare-earth metal are
calculated for $Q^{*}=1$.

$
\begin{tabular}{|l|l|l|l|l|l|}
\hline
R & Tb & Dy & Ho & Er & Tm \\ \hline
$J$ & 6 & 15/2 & 8 & 15/2 & 6 \\ \hline
$\alpha _J\times 100$ & -1.0 & -0.63 & -0.22 & 0.25 & 1.01 \\ \hline
$\langle r_f^2\rangle $ & 0.76 & 0.73 & 0.69 & 0.67 & 0.64 \\ \hline
$c/a$ & 1.58 & 1.57 & 1.57 & 1.57 & 1.57 \\ \hline
$K_1^{\exp }$ & -123 & -114 & -50 & 58 & 100 \\ \hline
$Q^{*}$ & 1.3 & 1.1 & 1.2 & 1.45 & 1 \\ \hline
$K_{1\text{loc}}^{\text{calc}}/Q^{*}$ & -2100 & --2050 & -800 & 800 & 2000
\\ \hline
RCo$_{5+x}$ & TbCo$_{5.1}$ & DyCo$_{5.2}$ & HoCo$_{5.5}$ & ErCo$_6$ &  \\ 
\hline
$K_1^{\exp }$ & -96 & -211 & -203 & 80 &  \\ \hline
$Q^{*}$ & -0.02 & -0.03 & -0.09 & -0.04 &  \\ \hline
\end{tabular}
$

Another interesting example is the case of imbedded positive muons\cite
{kron,Hof} which induce a strong local anisotropy. This leads to strong
deformation of the magnetic structure near the muon and to a considerable
contribution to the dipole field at the point of its location\cite{Cam,ii}.
According to Table 1, for $Q^{*}\sim 1$ the local anisotropy constants make
up about 10$^3$K and are large in comparison with the values of the
molecular field acting on the rare-earth ions, $\lambda \sim 100$K.
Therefore a strong cast of magnetic moments should take place in the limit
of large $K_{1\text{loc}}$. In the case $K_1>0,$ $K_{1\text{loc}}>0$ for the
ferromagnetic ordering along the $z$-axis we have for the dipole field 
\begin{equation}
\Delta B_{\text{dip}}^z=4\sqrt{3}M/a^3\sim 25\,\text{kG}
\end{equation}
where $M\simeq \sqrt{J(J+1)}\mu _B$ is the magnetic moment of rare-earth
ion. In the case $K_1<0,$ $K_{1\text{loc}}<0$ the direction and value of the
dipole field depend on the relation between $K_1$ and $\lambda $ (see Ref.%
\cite{ii}), but $|\Delta {\bf B}_{\text{dip}}|$ has the same order of
magnitude. Thus the local distortion of the magnetic structure yields the
contribution to $\Delta {\bf B}_{\text{dip}}$, which is of the same order as
the sum over the regular lattice (cf. the calculation for holmium \cite{Nik}%
; note that the nearest-neighbor ions give zero contribution to this sum in
the absence of the distortion). Phenomena which occur in ferromagnetic and
helical phases at intermediate values of $K_{1\text{loc}}$ with changing the
relation between $K_{1\text{loc}},\,K_1$ and $\lambda $ (e.g., with the
change of temperature or external magnetic field) are discussed in Ref.\cite
{ii}.

To conclude, the effects of screening of ion (impurity) charge by conduction
electrons turn out to be very strong: the anisotropy constants can be
strongly modified and even change their signs. More quantitative
investigations with account of a real electronic structure, multi-center
effects and non-spherical charge-density distribution seem to be important
for the problem of magnetic anisotropy in the systems under consideration.

The research described was supported in part by the Grant N 96-02-16999-a
from the Russian Basic Research Foundation.

{\sc Figure captions}

Fig.1. The distance dependence of the function $4\pi R^3\delta Z(R)$ for $%
k_Fd=2$ (solid line) and $k_Fd=3$ (dashed line).

Fig.2. The dependence $4\pi R^4\delta Z^{\prime }(R)$ for the same parameter
values as in Fig.1.

Fig.3. The distance dependence of the effective charge $Q^{*}(R)$ for the
same parameter values as in Fig.1.

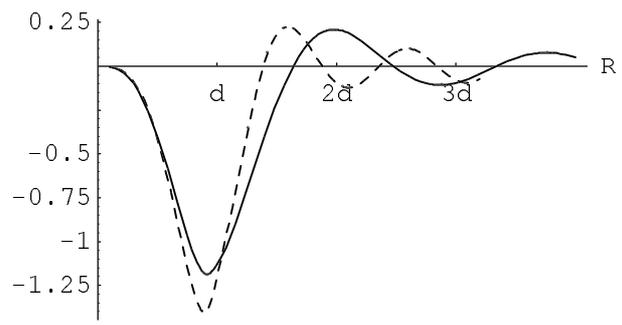
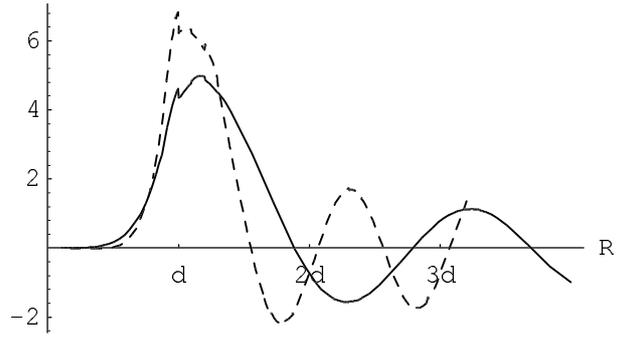
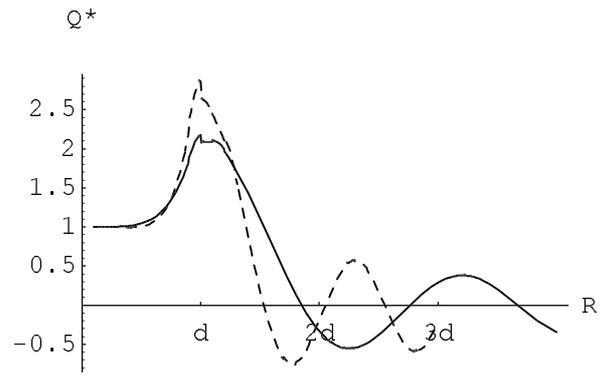
\end{document}